\documentclass[a4paper,10pt]{article}
\usepackage{graphicx}
\usepackage{latexsym,amsmath,verbatim}

\newcommand{\bea}{\begin{eqnarray}}
\newcommand{\eea}{\end{eqnarray}}

\title{Minimal Agent Based Model for Financial Markets I: Origin and Self-Organization of Stylized Facts}

\author{V. Alfi$^{1,2}$, M. Cristelli$^1$, L.Pietronero$^{1,3}$, A. Zaccaria$^1$\\\\
$^1$ \small Universit\`a ``La Sapienza'', P.le A. Moro 2, 00185, Roma, Italy\\
$^2$ \small Centro ``E. Fermi'', Compendio Viminale, 00184, Roma, Italy\\
$^3$ \small  ISC-CNR, V. dei Taurini 19, 00185, Roma, Italy
}

\begin{document}

\maketitle

\begin{abstract}
We introduce a minimal Agent Based Model for financial markets 
to understand the nature and Self-Organization of the Stylized Facts.
The model is minimal in the sense that we try to identify the essential  
ingredients to reproduce the main most important deviations
of price time series from a Random Walk behavior.
We focus on four essential ingredients: fundamentalist agents
which tend to stabilize the market; chartist
agents which induce destabilization; analysis of price behavior
for the two strategies; herding behavior which governs the possibility of
changing strategy.
Bubbles and crashes correspond
to situations dominated by chartists, while fundamentalists provide a long time
stability (on average).
The Stylized Facts are shown to correspond to an intermittent behavior  
which occurs only for a finite value of the number of agents $N$.
Therefore they correspond to finite size effect which,
however, can occur at different time scales. We propose a new mechanism for the 
Self-Organization of this state which is linked to the existence
of a threshold for the agents to be active or not active.
The feedback between price fluctuations and number of active agents 
represent a crucial element for this state of Self-Organized-Intermittency.
The model can be easily generalized to consider more realistic variants.

\end{abstract}

\section{Introduction}

In the past years there has been a large interest in the development 
of Agent Based Models (ABM) aimed at reproducing and understanding the Stylized
Facts (SF) observed in the financial time series. 
The simplest model of price time series is the Random Walk (RW)
introduced by Louis Bachelier in 1900~\cite{bachelier}.
The availability of large amounts of data has revealed a variety of
systematic deviations from the RW which are the SF and are
relatively common to all markets.
As we are going to see the properties of the SF resemble Critical Phenomena
and Complex Systems with possible relations to various problems in Statistical 
Physics. Many questions are however open and also how far one can go 
with these similarities represents a highly debated problem.
\\
The main SF, discussed in detailed in~\cite{ramac,mantegnabook,bouchaud} are:
\\
\\
{\bfseries{Absence of linear auto-correlations}}: It is not possible to 
predict the direction of the next price change from the previous one. 
Only at very short times (minutes or less) some correlations, 
positive or negative, 
can be present for technical reasons. This property corresponds to the 
condition of an efficient market (no simple arbitrage), which implies 
that it is not possible to make a profit without risk. 
If such a possibility would 
occur the market would immediately react and eliminate it. 
\\
{\bfseries{Heavy Tails}}: The distribution of returns (price increments) 
has a shape which 
is markedly different from the Gaussian case (RW). 
The probability for large positive or negative fluctuations is much 
larger than for a Gaussian distribution. These fat tails have been 
fitted in various ways and in some range they can be approximated by a power 
law with an exponent ranging from 3 to 5.
This behavior is observed for time scales ranging from minutes to 
weeks, while for very long times, a Gaussian distribution seems to be recovered.
\\
{\bfseries{Volatility Clustering}}: The variance of the returns shows a 
long range correlation in time. 
According to Mandelbrot~\cite{mandel,mande2}: ``large changes tend to be 
followed by large changes, of either sign, and small changes tend to 
be followed by small changes''. The correlation of the absolute values 
of the square of the returns is usually fitted by a power law  
(with exponents ranging from about zero to 1.0 and most probable value around
0.3) which can extend from minutes to weeks.
\\
These three main SF have been interpreted in term of various phenomenological 
stochastic models which permit in some cases an estimate of the risk 
beyond the Black and Scholes model which corresponds to the simple RW
~\cite{mande2,engle,tsallis2003,borland,stella}. 
\\
\\
In our opinion it would be important to add the following additional elements
which are also essential to built a conceptual framework necessary to 
understand market dynamics.
\\
{\bfseries{Nonstationarity and Time Scales}}: A characteristic of the 
price dynamics is that it is not clear if one can define a probability 
distribution which is stationary in time and, in any case, 
it is not easy to estimate the appropriate time scale~\cite{lebaronbreve}. 
This is a crucial point that should be carefully analyzed in any statistical 
treatment of the data.
\\
{\bfseries{Self-Organization}}:
A point which is usually neglected 
is the fact that the stylized facts usually correspond to a particular situation
of the market. If the market is pushed outside such a situation, it will
evolve to restore it spontaneously.
The question is why and how all markets seem to self-organize in this 
special state.
The answer to this question represents a fundamental point in understanding
the origin of the SF.
\\
\\
In this paper we discuss a minimal ABM~\cite{paperNP}
which includes the following elements (first considered by Lux and Marchesi in~\cite{LMnature,LM}):
\begin{itemize}
\item fundamentalists: these agents have as reference a fundamental price $p_f$
derived from standard economic analysis of the value of the stock.
Their strategy is to trade on the fluctuations from this reference value
and bet on the fact that the price will finally converge towards the 
reference value.
These traders are mostly institutional and their time scale is
relatively long.
Their effect is a tendency to stabilize the price around the reference value.
\item chartists: they consider only the price time series and
tend to follow trends in the positive or negative direction.
In this respect they induce a destabilizing tendency in the market.
These traders have usually a time horizon shorter than the fundamentalists
and they are responsible for the large price fluctuations corresponding
to bubbles or crashes.
\item herding effect: this is the tendency to follow the strategy
of the other traders. It should be noted that  traders can change their 
strategy from fundamentalist to chartist and vice-versa depending 
on various elements.
\item price behavior: each trader looks at the price from her perspective
and derives a signal from the price behavior. This signal will be
crucial in deciding her trading strategy.
\end{itemize}
These four elements are, in our opinion, the essential irreducible
ones that an ABM should possess and we are going to describe
in detail their combined effects. Of course real market are extremely
more complex but the study of these elements represents a basis on which one
might eventually add more realistic features.
In this perspective we will try to construct a model which is the simplest
possible containing these four elements. 
This permits to obtain
a detailed understanding of the origin and nature of the SF
and also to discuss their self-organization.
The main result of this paper is that the SF correspond to finite size 
effect in the sense that they disappear in the limit of large $N$ 
(total number of agents) or large time. 
Given that the agents' strategies can be defined for different time scales
this finite size effects can act at different time and 
resemble power laws and critical exponents.
However, in our perspective, these exponents are essentially a fit to
the data without the important property of universality.
This situation has important implication both for the microscopic 
understanding of the SF as well as for the analysis and interpretation of 
experimental data.
The concept of self-organization arises very naturally from the introduction
of a threshold in the agents' activity.
Namely a price which is very stable (which corresponds to a very large $N$) 
demotivates agents to trade this 
stock and will naturally lead to a decrease of the number of agents.
On the other hand a small number of agents leads to large fluctuations
in the price which presents opportunities of arbitrage and this 
will attract more traders.
So, in the end, the system will self-organizes around the number of 
traders which corresponds to a situation of intermittency, leading to the SF.
This phenomenon does not correspond to self organized
criticality precisely but rather to a Self Organized Intermittency
(SOI) which occurs for a finite number of agents (on average).
\\
In Sec. 2 we define the basic elements of the minimal ABM in the perspective
of its maximum simplification.
\\
In Sec. 3 we consider in detail the fluctuations induced by the 
herding dynamics.
\\
In Sec. 4 we discuss the properties of the SF as arising from the model
for a specific range of parameters.
\\
In Sec. 5 we give an interpretation to the SF in terms of semi-analytical 
considerations.
\\
In Sec. 6 we discuss the self-organization of the market dynamics
towards the self-organized regime with the SF.
\\
In Sec. 7 we present our conclusions.
\\
In the next paper~\cite{paperoII}
we are going to focus on various statistical properties from
both a numerical and analytical point of view. 
We will also discuss in some detail the differences between linear 
and multiplicative dynamics.
In the present paper we only consider the linear dynamics for simplicity.

\section{Definition of a minimal ABM with Fundamentalists and Chartists}
\label{sec:2}
In this section we are going to describe in detail the ABM 
introduced in~\cite{paperNP}. 
The model is constructed to be as simple as possible but
still able to reproduce the SF of financial time series. 
This simplicity is very important because it  permits to derive
a microscopic explanation of the SF in 
term of the model parameters.
Elements which consider more realistic situations however can be added 
in a systematic way.
This model is inspired by the well known Lux-Marchesi (LM) 
model ~\cite{LMnature,LM}
but 
it is much simpler with respect both to the number of parameters  and
the rules for the dynamics.
\\
In our model we consider a population of $N$ interacting agents which are divided into two main categories: fundamentalists and chartists.
The fundamentalist agents tend to stabilize the market driving the price 
towards a sort of reference value
which we call the fundamental price ($p_f$). 
This kind of agents can be identified with long-term traders
and institutional traders~\cite{lillo-vaglica}. 
Chartists instead are short-term traders who look for detecting a local trend 
in the price fluctuation. This kind of agents tries to gain from detecting 
a trend and so they
are responsible for the formation of market bubbles and crashes which 
destabilize the market.
In the LM model the destabilizing effect made by chartists is implemented 
dividing the chartists agents
into two subcategories: optimists which always buy and pessimists which 
always sell.
In our model we can overcome this complication by simplifying the chartists' 
behavior.
In fact, the description of chartists is in term of the recently 
introduced potential method~\cite{taka1,taka2,vale2,vale3}.
Chartist agents detect a trend by looking at the distance between the 
price and its smoothed profile
which in our case is the moving average  
computed on the previous $M$ time steps.
Chartists try to follow the trend and bet that the price will moreover move 
away from the actual price.
In this way they create a local bubble which destabilizes the market.
The stochastic equation for the price which describes the chartists behavior 
can be written in terms of a RW
with a force whose center is the distance between the price and 
the moving average:
\begin{eqnarray}
p(t+1)&=&p(t)+\frac{b}{M-1}F(p(t)-p_M(t))+\sigma\xi(t)\\
\label{eq:0}
p_M(t)&=&\frac 1M \sum_{i=t}^M p(t)
\label{eq:1}
\end{eqnarray}
where $b$ is a parameter which gives the strength of the force, $p_M$ is the 
moving average performed on the previous $M$ steps of the price, $\xi$ is a white noise and $\sigma$ is the 
amplitude of this noise. 
The factor $M-1$ in the denominator of the force term makes the potential independent of the
particular choice of the moving average~\cite{taka1}. 
This equation has been used in previous papers to analyze 
real data~\cite{taka1,vale2,vale3}. The conclusion is that is indeed possible to observe this kind of 
forces also in real stock-prices. For our model we decided to adopt a simple linear expression for the force $(F=p(t)-p_M(t))$.

\begin{figure}[h!]
\centering
\includegraphics[angle=-90,scale=0.5]{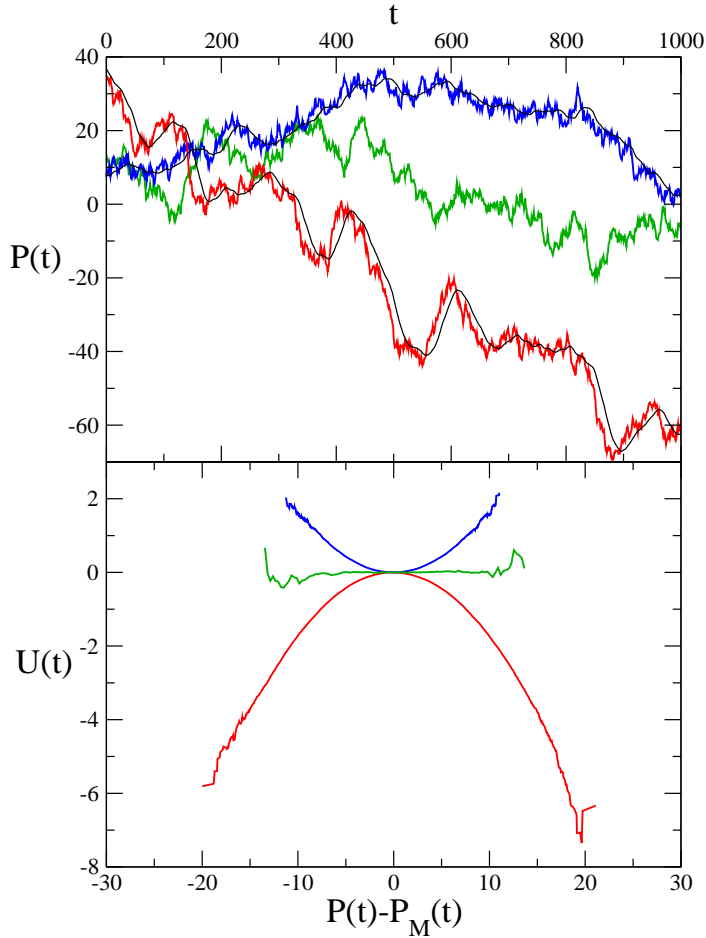}
\caption{The upper plot shows the price fluctuations for the process described by Eqs.(\ref{eq:1}) and (\ref{eq:0}) with a linear force $F$ for different values of the force strength parameter $b$. The values of the parameter $b$ are $b=-1$ (blue line), $b=0$ (green line) and $b=1$ (red line) which correspond respectively to attractive, flat and repulsive potentials as shown in the lower plot. In our model we are considering only the repulsive
potential which describes the behavior of the trend-follower chartists.}
\label{fig:1}
\end{figure}

By integrating the force it  is also possible to obtain the effective potential 
which describes the chartists' behavior. For the case of a simple linear 
force the potential is quadratic. In Fig.~\ref{fig:1} we show the stochastic 
process described in Eq. (\ref{eq:1}) with a linear force and the 
corresponding potential.
A random walk is also plotted for comparison.
Depending on the sign of the parameter $b$ one can obtain both attractive and 
repulsive potential. Here we are considering only the repulsive potential 
which describes the trend-follower behavior of chartists,
because the stabilizing attractive case is already carried out by the 
fundamentalist agents.
In fact, fundamentalists try to stabilize the market driving the price 
towards the fundamental price
$p_f$ which in our model is constant in time. 
The stochastic equation which describes the fundamentalists' behavior can be 
written in terms of a random walk with a further term which is responsible 
for the stabilizing action of fundamentalists:
\begin{equation}
p(t+1)=p(t)+\gamma(p_f-p(t))+\sigma\xi(t) 
\label{eq:2}
\end{equation}
where $\gamma$ is the strength of the fundamentalists' action.
For the moment the total number of agents, $N$, is kept fixed and there 
are $N_f$ fundamentalists and $N_c$ chartists but
agents can decide to change their mind during the simulation by switching their 
strategy from fundamentalist to chartist and viceversa. 
The probability to change strategy is based on two terms.
The first is an herding term: agents tend to imitate other people 
behavior proportionally
to the relative number of agents in the arrival class.
The second is a term, characterized by the parameter $k$, which leaves 
the agents the possibility to change 
strategy  on the basis of considerations on 
the price behavior, independently
on what the other agents are doing. 
The price signal which appears in the transition probabilities is 
proportional to 
$|p(t)-p_f|$ for the probability to become fundamentalist 
and $|p(t)-p_M(t)|$ for that to 
become chartists. 
The mathematical expression for the transition probabilities is:
\begin{eqnarray}
\label{eq:3}
P_{cf}&\propto& (K+\frac{N_f}{N})\exp{(\gamma |p(t)-p_f|)}\\ 
P_{fc}&\propto& (K+\frac{N_c}{N})\exp{(\frac{b}{M-1} |p(t)-p_M(t)|)}
\label{eq:4}
\end{eqnarray}
For a realistic representation of the market the fundamentalists should 
dominate for very long times with intermittent appearance of 
bubbles or crashes due 
to the chartists.
This is due to the fact that fundamentalists are usually 
big institutional traders
which have on average a major weight in the market. 
In order to properly reproduce this effect we can introduce
an asymmetry in the herding dynamics and, neglecting the price
modulation,
Eqs. (\ref{eq:3}) and (\ref{eq:4}) can be written in the following way:
\begin{eqnarray}
P_{cf}&\propto& (K+\frac{N_f}{N})(1+\delta)\\ 
\label{eq:5}
P_{fc}&\propto& (K+\frac{N_c}{N})(1-\delta)
\label{eq:6}
\end{eqnarray}
where the positive parameter $\delta$ define the asymmetry of the model.
In the case $\delta=0$ the model reduces to the symmetric ants model by Kirman~\cite{Kirman:1993}.
In Sec.~\ref{sec:3} we will analyze in detail this asymmetric model and the
corresponding equilibrium distribution for the population of chartists and fundamentalists.
For the price formation we are going to use a simple linearized form of the 
Walras' Law~\cite{kreps}:
\begin{equation}
p(t+1)-p(t)=\Delta p\propto ED(t)
\label{eq:7}
\end{equation}
where $ED$ is the excess demand of the market at time $t$. 
This linear interpretation, valid
for small price increments, ($\Delta p \ll 1$)  is a technical
simplification from the more realistic multiplicative dynamics and it is 
very useful for an analytical treatment.
The more general multiplicative
case will be discussed in detail in a forthcoming paper~\cite{paperoII}.
With respect to the dynamics of the ABM the multiplicative case will turn out 
to be less stable in terms of parameters while the the linear case is
mathematically less problematic.
\\
In our model the excess demand $ED$ is simply proportional to 
the price signal of
chartists and fundamentalists:
\begin{equation}
ED=ED_f+ED_c=\frac{N_f}{N}\gamma(p_f-p(t))+\frac{N_c}{N}\frac{b}{M-1}(p(t)-p_M(t))
\label{eq:8}
\end{equation}
then adding a noise term we have the complete equation for the price formation:
\begin{equation}
p(t+1)-p(t)=ED+\sigma\xi(t)
\label{eq:9}
\end{equation}
In this case we have that the volume of the agents' actions is proportional 
to the signal they perceive.
This situation is much simpler than the corresponding price 
formation for the Lux-Marchesi model
where prices can only vary of a fixed amount (tick) and the connection with 
the excess demand
is implemented in a probabilistic way.
\clearpage

\section{Herding Dynamics}
\label{sec:3}
\subsection{Symmetric case}

\begin{figure}
\centering
\includegraphics[scale=0.3]{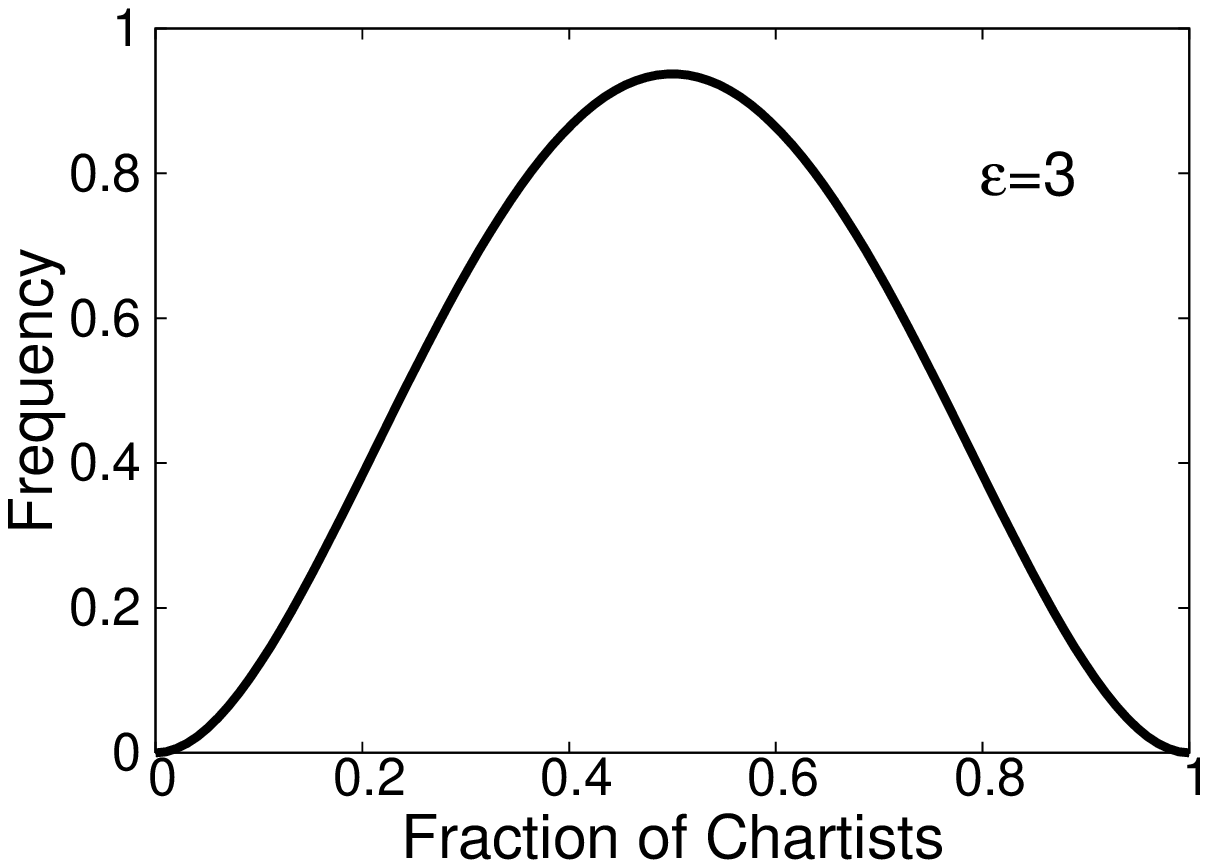}
\includegraphics[scale=0.3]{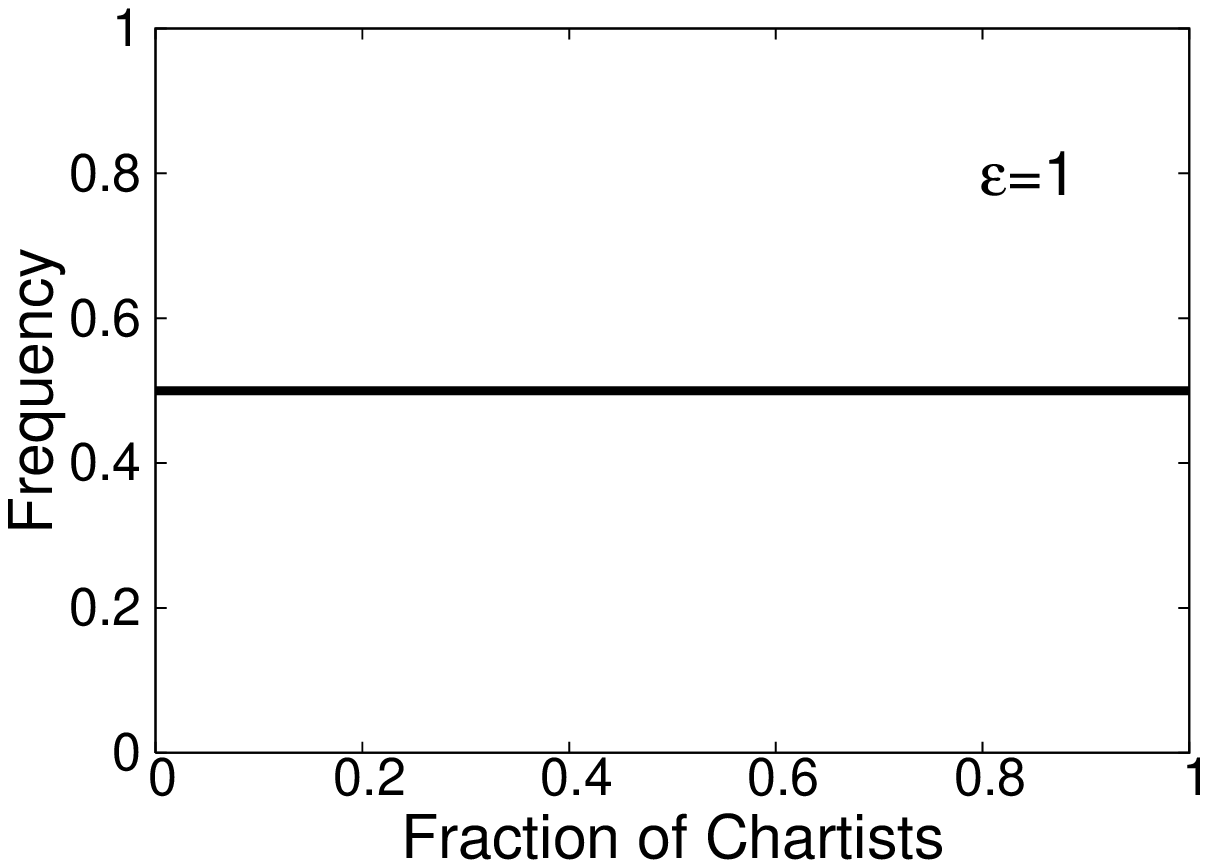}
\includegraphics[scale=0.3]{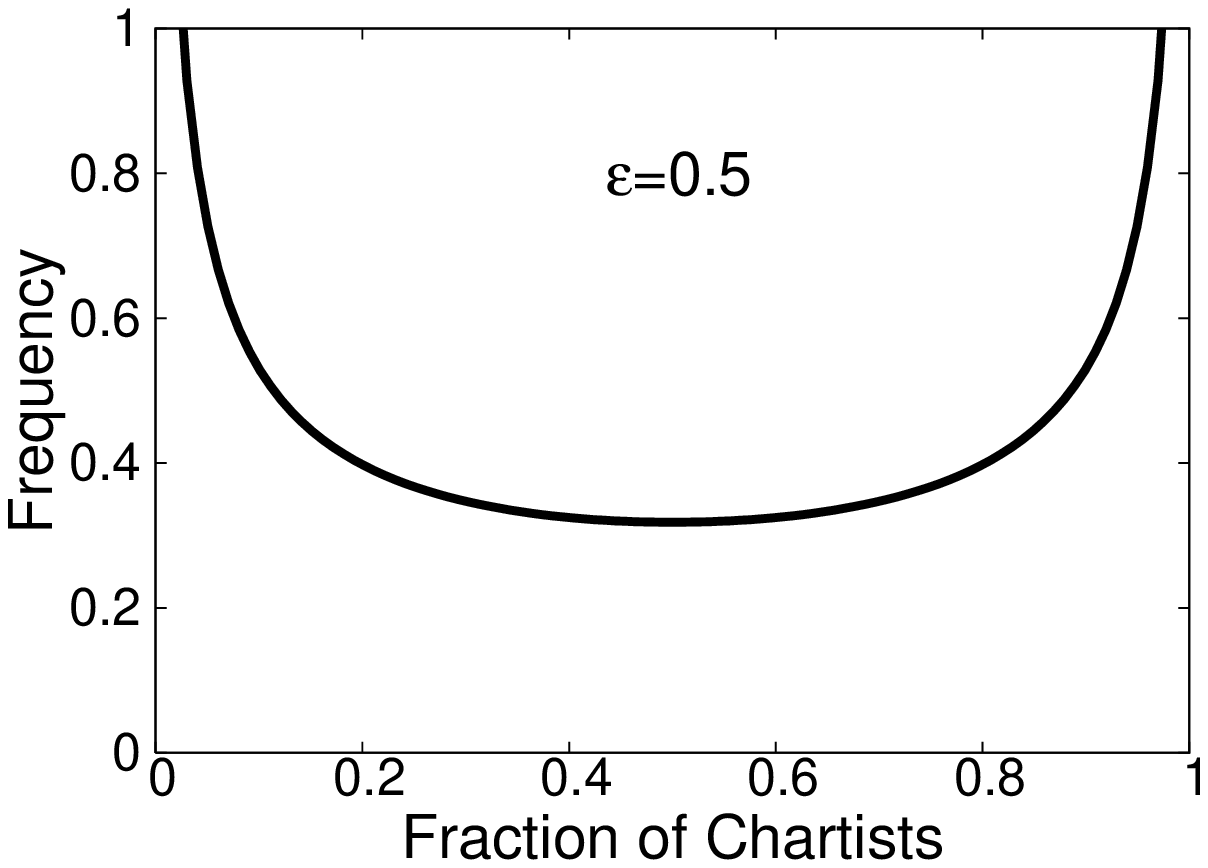}
\caption{The functional form of the distribution Eq. (\ref{equialtra}) 
changes with $\epsilon=KN$. We show the cases $\epsilon=3.0; 1.0$ and $0.5$. 
Only for $\epsilon<1$ the system shows an alternation between two metastable states.}
\label{dasaafsdfsw}
\end{figure}

In his seminal paper Kirman \cite{Kirman:1993} 
proposed a simple dynamical model to explain a peculiar behavior observed 
in ant colonies. Having two identical food sources, ants 
prefer alternatively only one of these, and (almost) 
periodically switch from one source to the other. In Kirman's 
model this effect is caused by an \textit{herding dynamics}, in 
which the evolution is stochastic and based on the meetings of the ants. 
In particular an ant can recruit a companion, and bring it 
to its preferred source, with a certain probability.\\
This model can be used to take into account the herding dynamics of 
a population of $N$ agents (with $N$ fixed), and it is 
formalized as follows.\\
Let us suppose the existence of two microscopic states: an agent 
at time $t$ can be either a \textit{fundamentalist} or 
a \textit{chartist}. Defining $N_c=N_c(t)$ as the number 
of chartists at time $t$ (and analogously $N_f=N_f(t)=N-N_c(t)$ as 
the the number of fundamentalists), the quantity
\begin{align}
x=\frac{N_c}{N}
\label{xdef}
\end{align}
varying between $0$ and $+1$ describes the macroscopic state of the system.\\
A possibility to change opinion is given to each agent at any given time step 
if she meets an agent with opposite views who succeeds in recruiting her. 
Under the hypothesis that the transition probabilities are small 
enough, in a single time step there will be no more than one change 
of opinion. In this case we can write the transition rates as
\bea
P_{cf}=\beta\frac{N_f}{N} \qquad P_{fc}=\beta\frac{N_c}{N}
\label{gyijhoiyujoi}
\eea
where $P_{cf}$ is the probability than \textit{one} agent passes from the chartist group
 to the fundamentalist one, and $\beta$ regulates the speed of the process.\\
Roughly, this is the approach adopted by Lux and Marchesi to rule the
transitions between agents in their model \cite{LM,LMnature}, apart an exponential term depending on agents' utilities that plays a minor
role in the dynamics. If we adopt the rates of Eq.(\ref{gyijhoiyujoi}) the dynamics
will admit two absorbing states ($N_c=0$ and $N_c=N$). Lux and Marchesi
assign a lower limit of 4 for each class of agents, while 
Alfarano et
al.~\cite{AL2}, following Kirman \cite{Kirman:1993}, suggested a Poissonian term 
(that is, independent from the number of agents) to avoid these fixed points. This new term can be 
seen as a spontaneous tendency to change one's mind, independently on the 
others. 
This corresponds essentially on the term $K$ we have discussed in relation
of Eqs. (\ref{eq:3}-\ref{eq:6}).
The simplest model is therefore given by the following transition probabilities.
We will adopt this last solution and write the transition probabilities as
\bea
 P_{cf}=\beta(K+\frac{N_f}{N}) \qquad P_{fc}=\beta(K+\frac{N_c}{N}).
\label{gyijhoi}
\eea
Simulating this process one can observe two different behaviors depending on
the choice of the parameters, and in particular on $\epsilon\equiv KN$. It can
be shown \cite{Kirman:1993} 
that if $\epsilon<1$ the system spends most of the time in the two 
\textit{metastable states} $x\approx 0$ and  $x\approx 1$, sometimes 
switching from one
to the other. In the case $\epsilon\geq1$, that is, if the self-conversion term is large enough, the systems fluctuates around $x=0.5$.\\
Kirman explicitly derived the functional form of the equilibrium function for $x$,
\bea
P_s(x)\sim x^{\epsilon-1}(1-x)^{\epsilon-1}.
\label{equialtra}
\eea
In Fig. \ref{dasaafsdfsw} the three cases $\epsilon>1$, $\epsilon=1$ and $\epsilon<1$ are shown.
In the first case the distribution has a single peek for $x=0.5$. By decreasing $\epsilon$ the distribution gets
smoother and smoother, until it becomes uniform for $\epsilon=1$. If one further decreases $\epsilon$, the distribution develops two peeks and becomes bimodal.\\

\subsection{Asymmetric case}

\begin{figure}[!ht]
\centering
\includegraphics[scale=0.3]{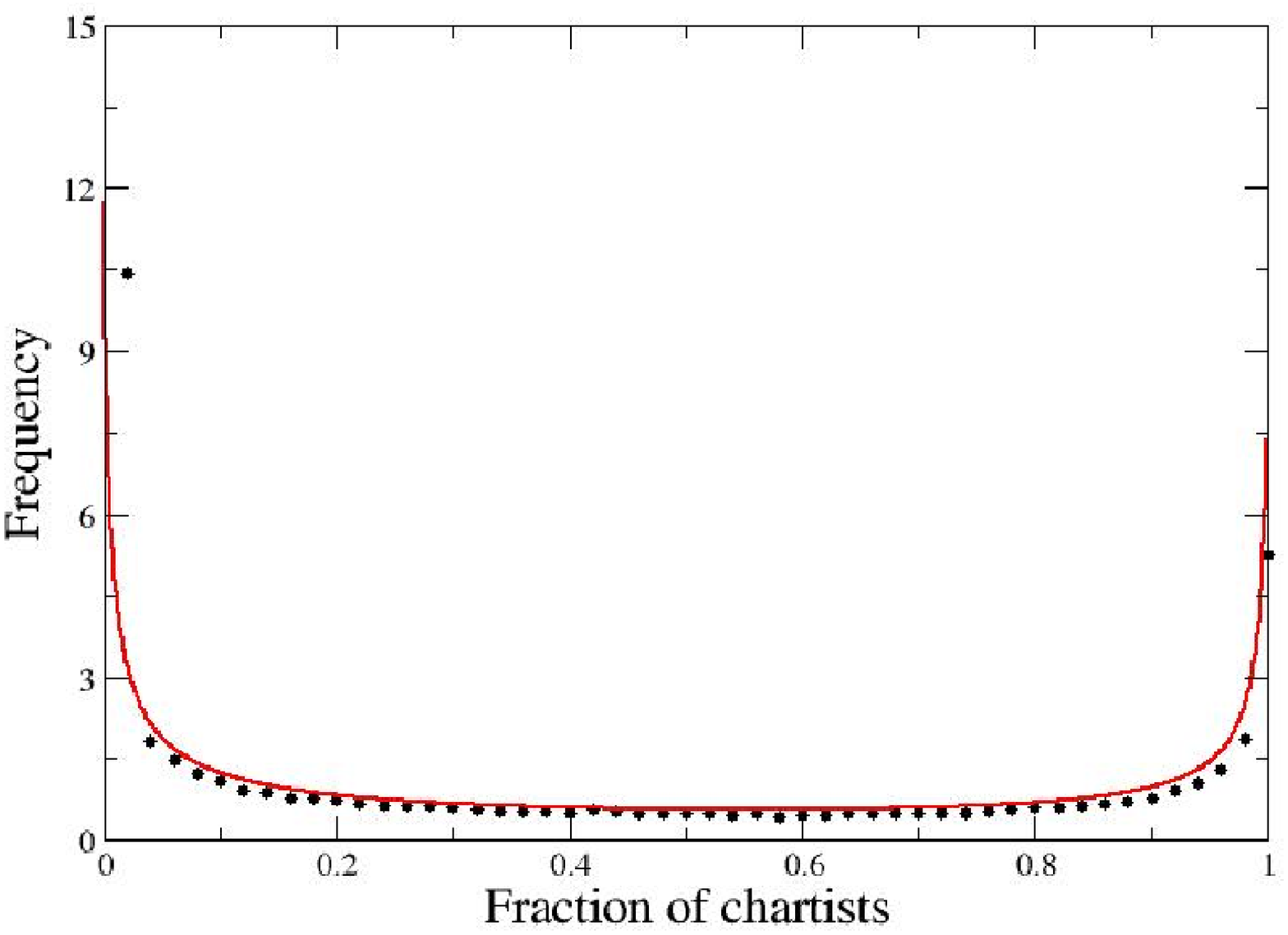}
\includegraphics[scale=0.3]{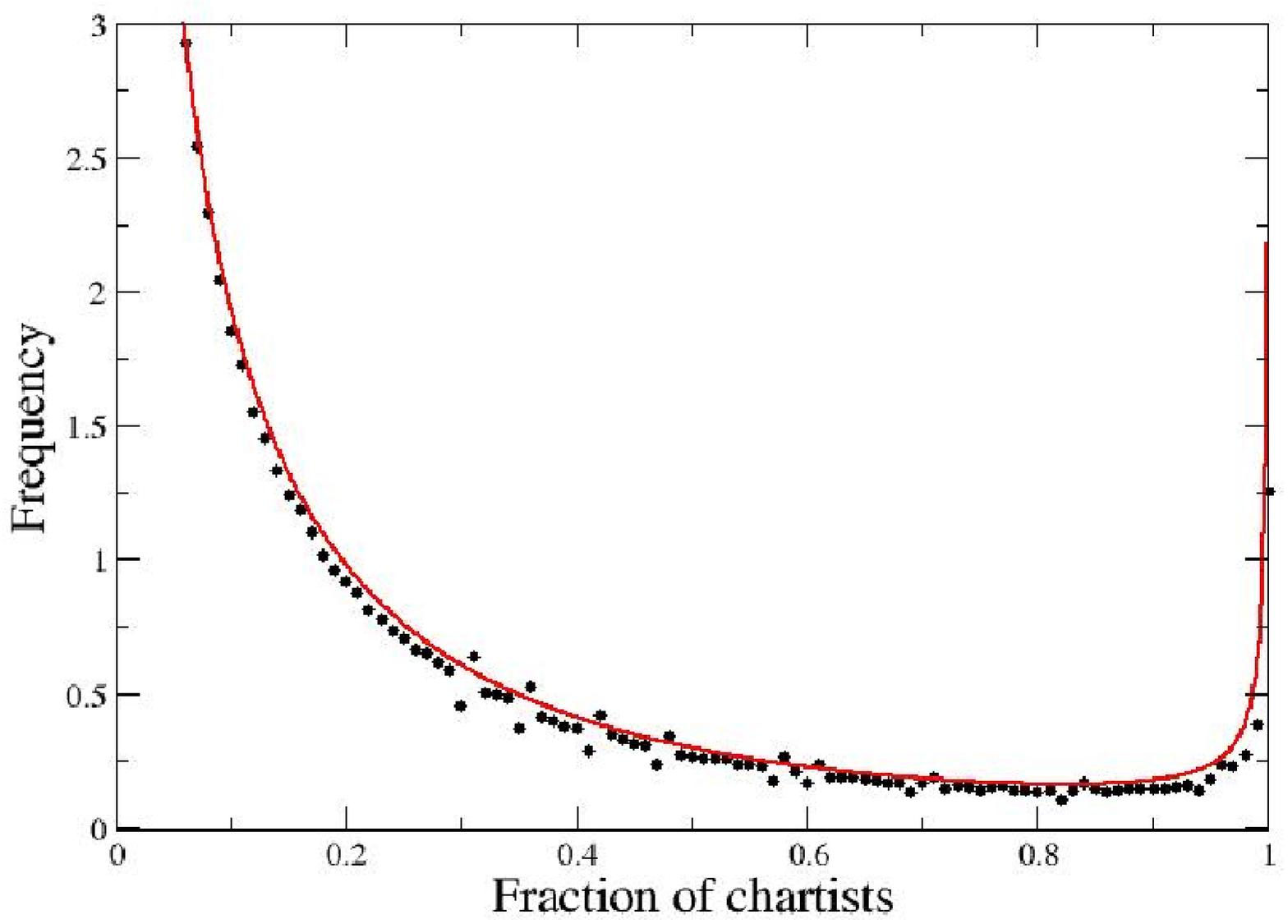}
\includegraphics[scale=0.3]{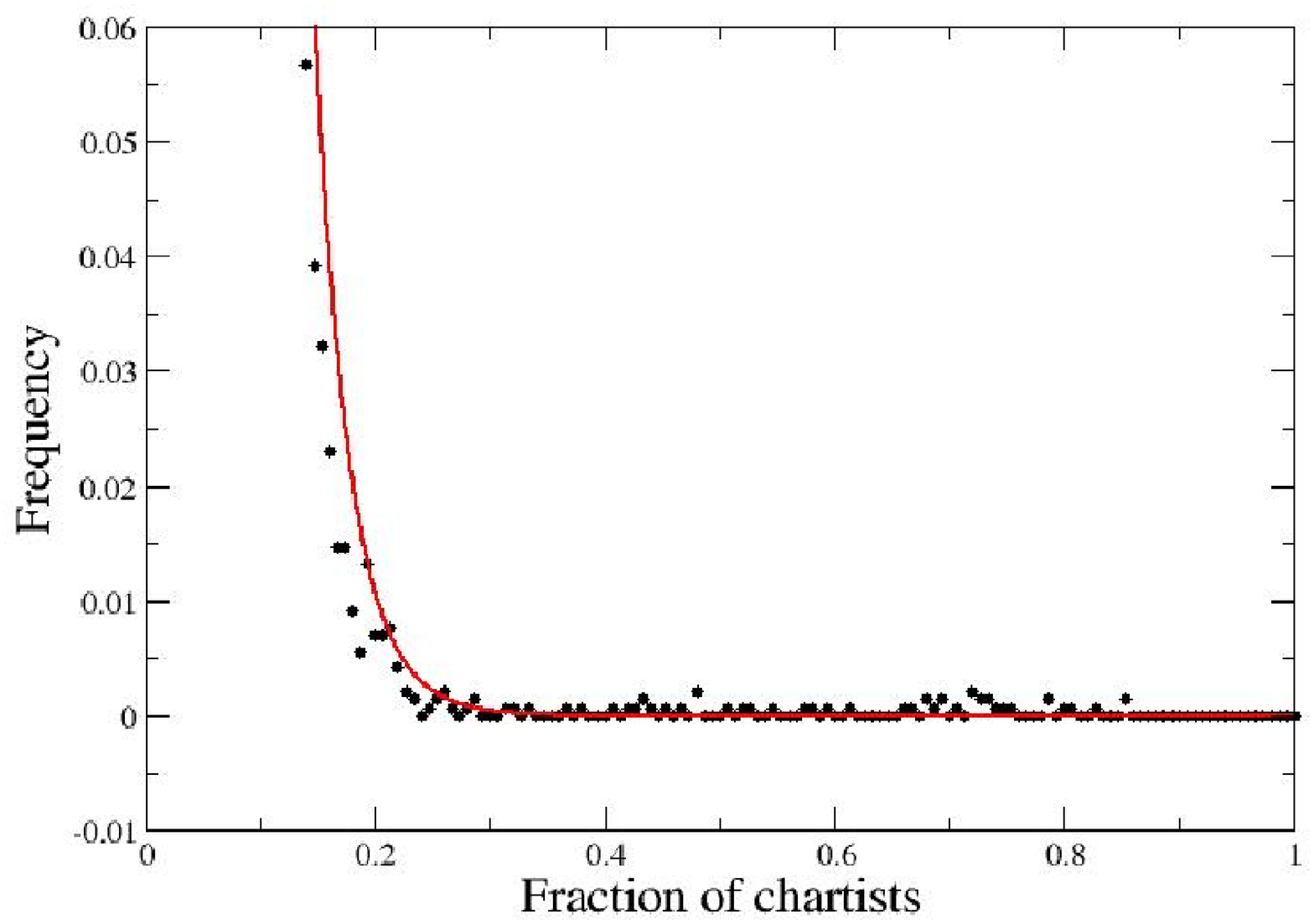}
\caption{Comparison between theory (red line) and simulation (black circles) for the asymmetric model described by the transition rates given by Eq. (\ref{sfhsdghsdgdgdgdg}). From top to bottom the three cases $N=50$, $N=500$ and $N=5000$ are shown, the other relevant parameter $\delta$ being fixed and equal to 0.003. The variation of the vertical scale underlines the progressive compression of the distribution towards the state in which fundamentalists are dominant.}
\label{fdfgdghshsh}
\end{figure}

It is possible to generalize the symmetric herding model described above to the
asymmetric case, in order to take account of the fact that institutional
traders, that have more impact on the market than individuals, usually adopt
long-term strategies~\cite{lillo-vaglica} , as fundamentalists do. Here we propose a
further generalization of the model proposed of Alfarano et al. \cite{alfasi},
which is more convenient for our purposes.\\
Our asymmetric dynamics is given by:
\[
p_{f \rightarrow c}=\beta (1-\delta)\big[K_1+\frac{N_c}{N}\big]
\]
\bea
p_{c \rightarrow f}=\beta (1+\delta)\big[K_2+\frac{N_f}{N}\big]
\label{sfhsdghsdgdgdgdg}
\eea
where $\delta>0$ regulates the asymmetry between the two metastable states
(the model described in \cite{alfasi} can be recovered in the case $\delta=0$).\\ 
We are going to see that the asymmetry introduced by the term $\delta$
is more realistic than the one due to $K_1$ and $K_2$ with respect to the
objective of having the market dominated (on average) by
fundamentalists at very large times.\\
It is possible to find the Fokker-Planck equation associated to the process given by Eq.(\ref{sfhsdghsdgdgdgdg}). Following Alfarano et al.~\cite{alfasi}, we consider
the Master Equation in the case in which, on average, we have no more than a change of opinion in a single time step (this approximation is valid if $\beta \ll 1$):
\[
 \frac{\Delta P_{N_c}(t)}{\Delta t}=P_{N_c+1} \pi(N_c+1 \rightarrow N_c)-P_{N_c} \pi(N_c \rightarrow N_c-1)
\]
\bea
-P_{N_c} \pi(N_c \rightarrow N_c+1)+P_{N_c-1} \pi(N_c-1 \rightarrow N_c).
\label{hihfivhdfivh}
\eea
where $P_{N_c}(t)$ is the probability to have $N_c$ chartists at time $t$ and $\pi$ is the rate of the transition in the brackets. In the continuous limit we can substitute the transition rates in Eq.(\ref{hihfivhdfivh}), finding after some passages the Fokker-Planck equation
\bea
\frac{\partial p(x,t)}{\partial t}= - \frac{\partial}{\partial x}A(x)p(x,t)+\frac{1}{2} \frac{\partial^2}{\partial x^2}D(x)p(x,t)
\eea
with drift and diffusion function given respectively by
\[
A(x)=\beta \bigg[-2\delta x (1-x)+(1-\delta)K_1(1-x)-(1+\delta)K_2 x-\delta\frac{1}{N^2}\bigg]
\]
and
\[
D(x)=\beta \bigg[\frac{2}{N}x(1-x)+\frac{(1-\delta)}{N}K_1(1-x)+\frac{(1+\delta)}{N}K_2 x-2\frac{\delta}{N^2}(x-\frac{1}{2})\bigg].
\]
Given these functions, the asymptotic stable distribution for the fraction $x$
of chartists is given by the textbook formula 
\bea
p_s(x,N)=\frac{C}{D(x)} \exp \bigg[ \int^x \frac{2A(y)}{D(y)} dy \bigg]
\label{dfhghighghgg}
\eea
where $C$ is a normalization constant (this expression can be obtained setting the temporal derivative of $p_s(x,N)$ equal to zero and integrating twice the Fokker-Planck equation, see for example \cite{Gard}).
\\
The explicit expression of the equilibrium distribution can be derived
by appropriate analytical computer codes and it is rather complex. 
Nevertheless a simple approximation will allow us to derive the distribution 
$\tilde{p}_s(x,N)$ which can be easily calculated and it illustrates clearly 
the role played by the asymmetry parameter $\delta$.\\
The idea is to disregard the terms of order $1/N^2$ in the drift and in the 
diffusion functions. This approximation is valid for $N\gg 1$ and, since we 
will deal with at least $\sim 50$ agents, we can expect that it is rather
appropriate to our case.
\\
Moreover, we are interested only in the case  $K_1N=K_2N<1$ in order to 
avoid the unimodality of the asymptotic distribution. This can be obtained setting $K_1=K_2=r/N$, with $r<1$. 
The reasons for adopting a parametrization in which $K$ is inversely
proportional to $N$ are:\\
(i) qualitatively one may expect that the probability to neglect the 
herding behavior decreases when $N$ increases.\\
(ii) in the following we will consider the properties of thee model as a 
function of $N$ and we believe it is realistic that the system always 
stays in the bimodal case. This requires that $\epsilon=KN$ should not 
exceed the value one even for large values of $N$.\\
With this choice and using the approximation described above, the drift and the diffusion functions become respectively 
 \bea
 \tilde A(x)=\beta \bigg[-2\delta x (1-x)+\frac{r}{N}\bigg(1-\delta-2x\bigg)\bigg]
 \eea 
 and
 \bea
 \tilde D(x)=\beta \bigg[\frac{2}{N}x(1-x)\bigg]
 \eea
Now we can easily solve the integral in Eq.(\ref{dfhghighghgg}) and derive
 \bea
  \tilde{p}_s(x,N) \propto x^{r(1-\delta)-1}(1-x)^{r(1+\delta)-1}\exp(-2\delta N x).
 \label{fhhdhlelee}
 \eea
The form of Eq. (\ref{fhhdhlelee}) clearly shows that, by increasing $N$, the 
values of $x$ near to 1 are exponentially suppressed. In other words, if $N$ 
is small the equilibrium distribution will be bimodal, because 
$\delta\ll1$ and $r<1$ (compare the Eqs.~(\ref{fhhdhlelee}) 
and (\ref{equialtra})), while by increasing $N$ the asymmetry, regulated by 
the parameter $\delta$, becomes more and more important.\\
We have also simulated the process described by the 
Eq. (\ref{sfhsdghsdgdgdgdg}) and we have compared the distributions obtained by integrating the Eq. (\ref{dfhghighghgg}) with the complete 
drift and diffusion functions. As it is clear from Fig. \ref{fdfgdghshsh}, 
the agreement between theory and simulation is very good.

\subsection{Finite size effects}

\begin{figure}[!ht]
\centering
\includegraphics[scale=0.3]{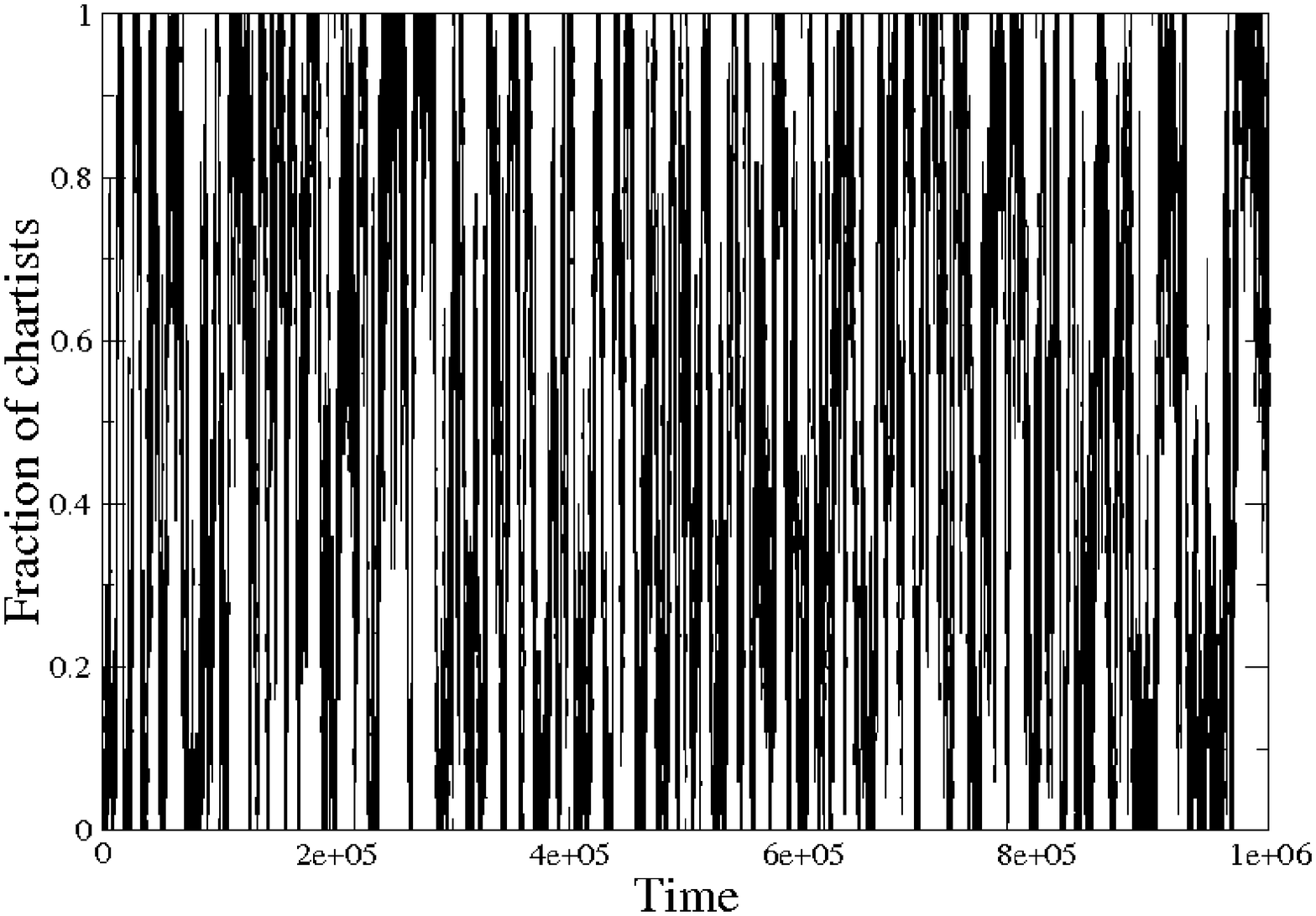}
\includegraphics[scale=0.3]{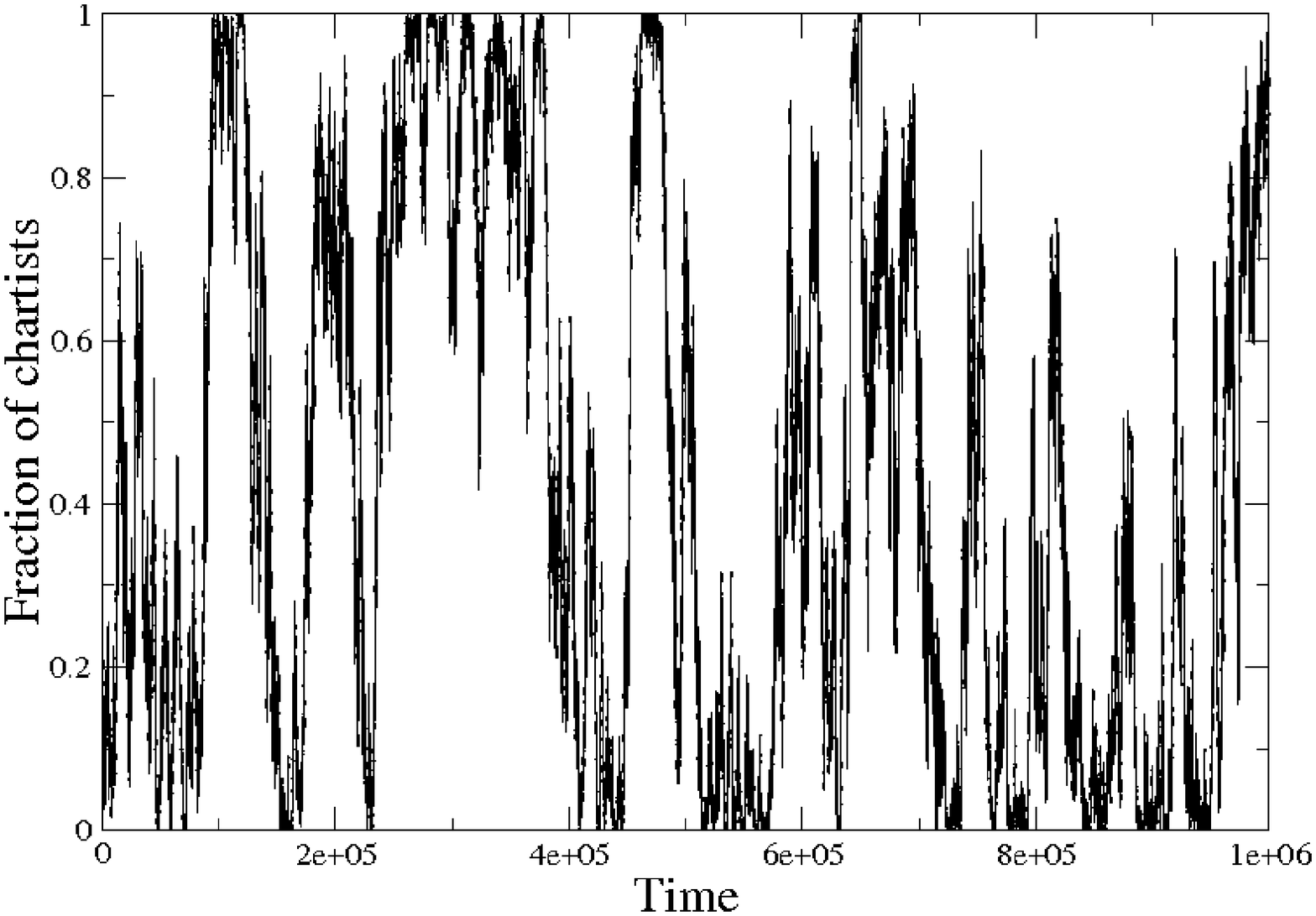}
\includegraphics[scale=0.3]{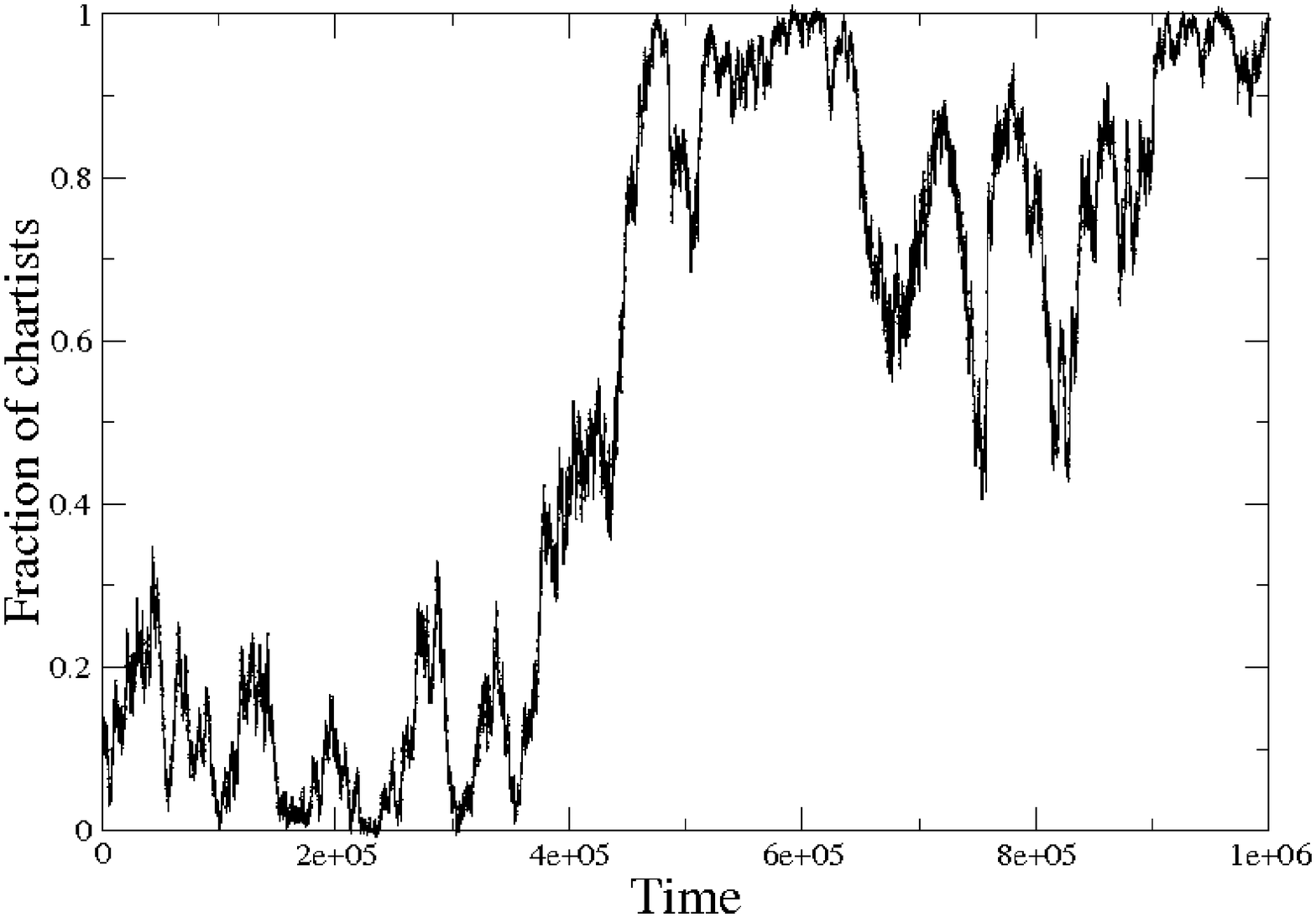}
\caption{Fraction of chartists $x$ for the dynamics given by 
Eq.(\ref{gyijhoiyujoi}). We have considered different values of $N$ but 
the same temporal scale for the horizontal axis, $T=10^6$ time steps. 
From top to bottom we have $N=50$, 500 and 5000 respectively. 
One can see that the cases $N=50$ and $N=5000$ are unrealistic, corresponding 
to too many or too few fluctuations. Only the case $N=500$ shows the 
right intermittency which leads to Stylized Facts.}
\label{flsgfd}
\end{figure}

Alfarano et al. \cite{AL2} 
computed the time that on average must be waited to see the switch from one metastable state to the other in the symmetric case:
\[
T_0=\frac{N}{\beta}\frac{\pi}{(1-2\epsilon)}\frac{\cos(\pi\epsilon)}{\sin(\pi\epsilon)}.
\]
The presence of this characteristic temporal scale implies that this simple 
model would give exponential correlation functions for volatility clustering. 
However different time scales in agents' strategies can  lead to a 
superposition of different characteristic times
and therefore to long-range relaxation.\\
Examining the functional form of $T_0$ one can see that for 
$\epsilon\rightarrow0$ the mean first passage time diverges, because of the 
emergence of the two absorbing states. Moreover, keeping $\epsilon$ constant 
and increasing $N$ the time spent in one of the metastable states gets longer.\\
We have simulated the process given by Eq.(\ref{gyijhoiyujoi}) using different 
values of $N$ keeping $\epsilon$ fixed and equal to 0.5. For the velocity 
parameter $\beta$ we choose the value 0.02.
As expected, the system oscillates between the two metastable states. 
As shown in Fig. \ref{flsgfd}, the only difference is the frequency of the 
passages from one state to the other.
For small values of $N$ ($N=50$) the rate to change strategy is very
high and this leads to an unrealistic situation in which fluctuations
are too fast. On the other hand, for large values of $N$ $(N=5000)$
the system get essentially locked in one of the two states and the 
fluctuations become frozen.
Only for an intermediate values of $N$ ($N=500$)
the system shows an intermittent behavior which resembles
experimental observations and will lead to the SF.
This finite size effect was first noted in~\ref{LMfinite}
in relation to the LM model.
In conclusion, this simple model explains why, in the limit 
$N\rightarrow\infty$, models like the Lux-Marchesi one loose their 
property to generate the SF.\\
This result is based on very general properties
and it is easy to expect that it will be a general one
for the entire class of model we are considering.
This implies that in this kind  of models, given a fixed 
temporal scale, the system 
will be able to generate the ``right'' 
amount of fluctuations only within 
a specific range of finite values of $N$.
This is a very important observation from both
a conceptual and a practical point of view.
In fact in this class of models the quasi-critical state
linked to the SF, corresponds to a finite size effect
(with respect to $N$ and to $t$) in the sense of Statistical Physics.
Therefore we are not in presence of a real critical behavior
characterized by universal power law exponents.
The fact that in some cases one can fit the experimental data with power laws
can be easily understood by considering that different agents might operate at
different time scales.
This would lead to a superpositions of finite size effects corresponding 
to different
time scale switch may appear as a sort of effective power law exponent.
The possibility that a suitable coupling between different time scale exist
leading to genuine critical behavior is of course open.
However, this is not the case for the class 
of model we are considering.
In this perspective the variability of the effective exponents
and their breakdown observed in various data~\cite{ramac}
can be a genuine effect not simply due to limitation
or problems with a database.
This of course change the perspective of the data analysis
and of their comparison with the models.
We expect that the behavior of different  markets is reasonably 
similar because the key elements are essentially the same
but without a strict universality.
\\
In the asymmetric case one must consider two different temporal scales, 
say $T_1$ and $T_2$, the first referring to the formation of the chartists' 
bubble and the second relative to its duration. We have investigated with 
numerical simulations their dependence on $N$ and $\delta$, finding, as 
expected, a divergence of $T_1$ in the limit $N\rightarrow\infty$.
Therefore the introduction of an asymmetry in the
agents dynamics does not change the finite size effect
associated to the quasi-critical behavior and the SF.

\clearpage
\section{Stylized Facts from the Model}

\begin{figure}[h!]
\centering
\includegraphics[angle=-90,scale=0.5]{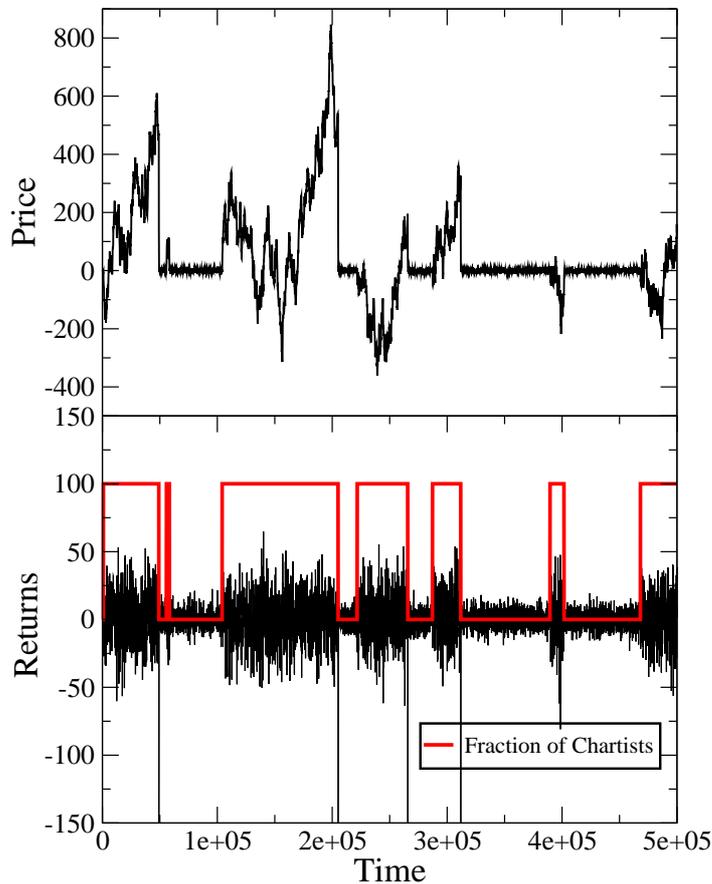}
\caption{Even a one agent model subjected to the stochastic dynamics
can lead to non trivial results in terms of fat tails and volatility clustering.
Here we can see the corresponding price and returns fluctuations.
In this case we have also simplified the probability to
change strategy by neglecting the exponential term
corresponding to the price behavior. In the figure
below the large returns are clearly related to the periods in which
the agent is a chartist. The price behavior may look unrealistic
because we consider a case with a fixed value of $p_f$. If one
would also introduce a random walk behavior for $p_f$ the price
behavior would look much more realistic.}
\label{fig:2}
\end{figure}

\begin{figure}[h!]
\centering
\includegraphics[angle=-90,scale=0.4]{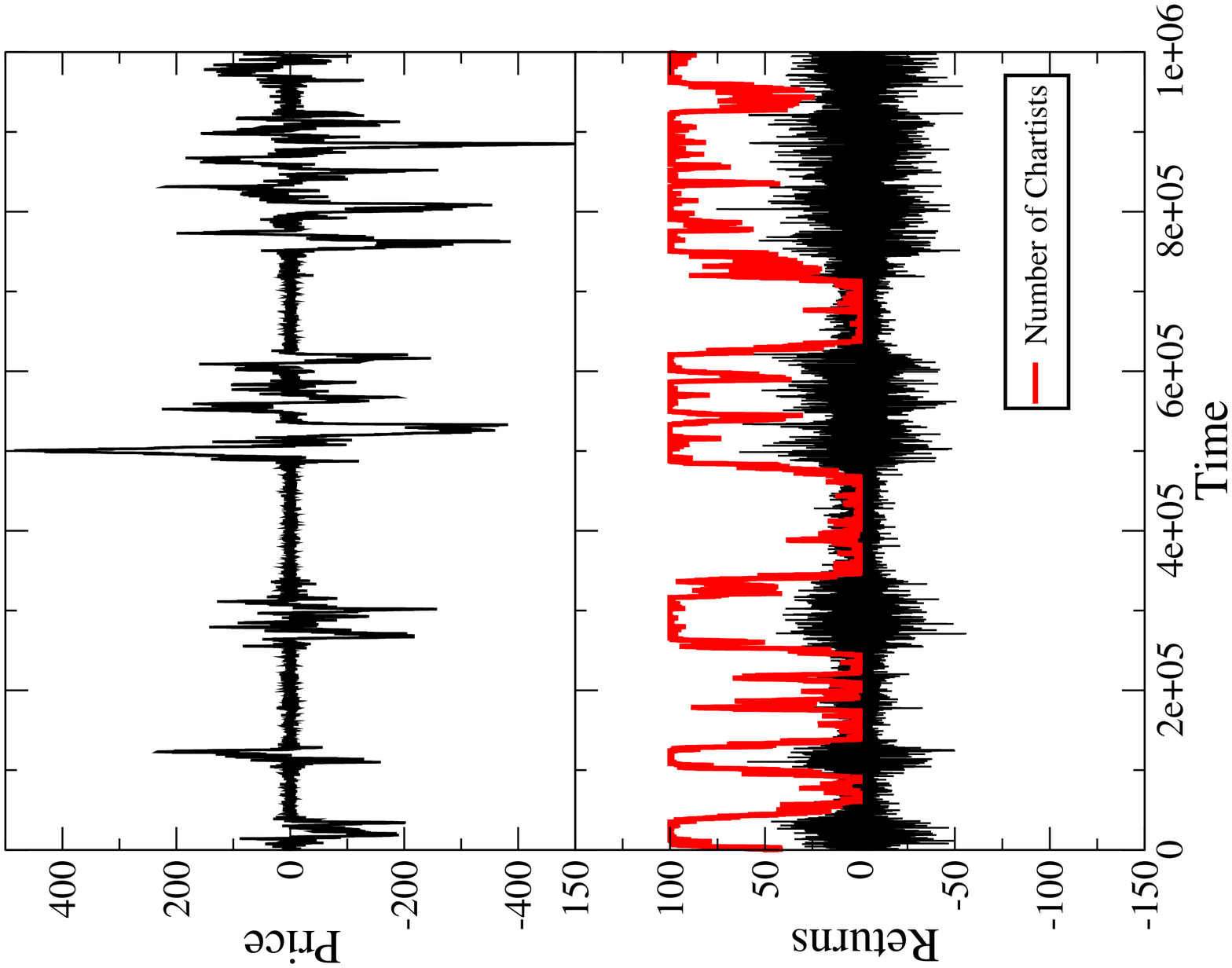}
\includegraphics[scale=0.4]{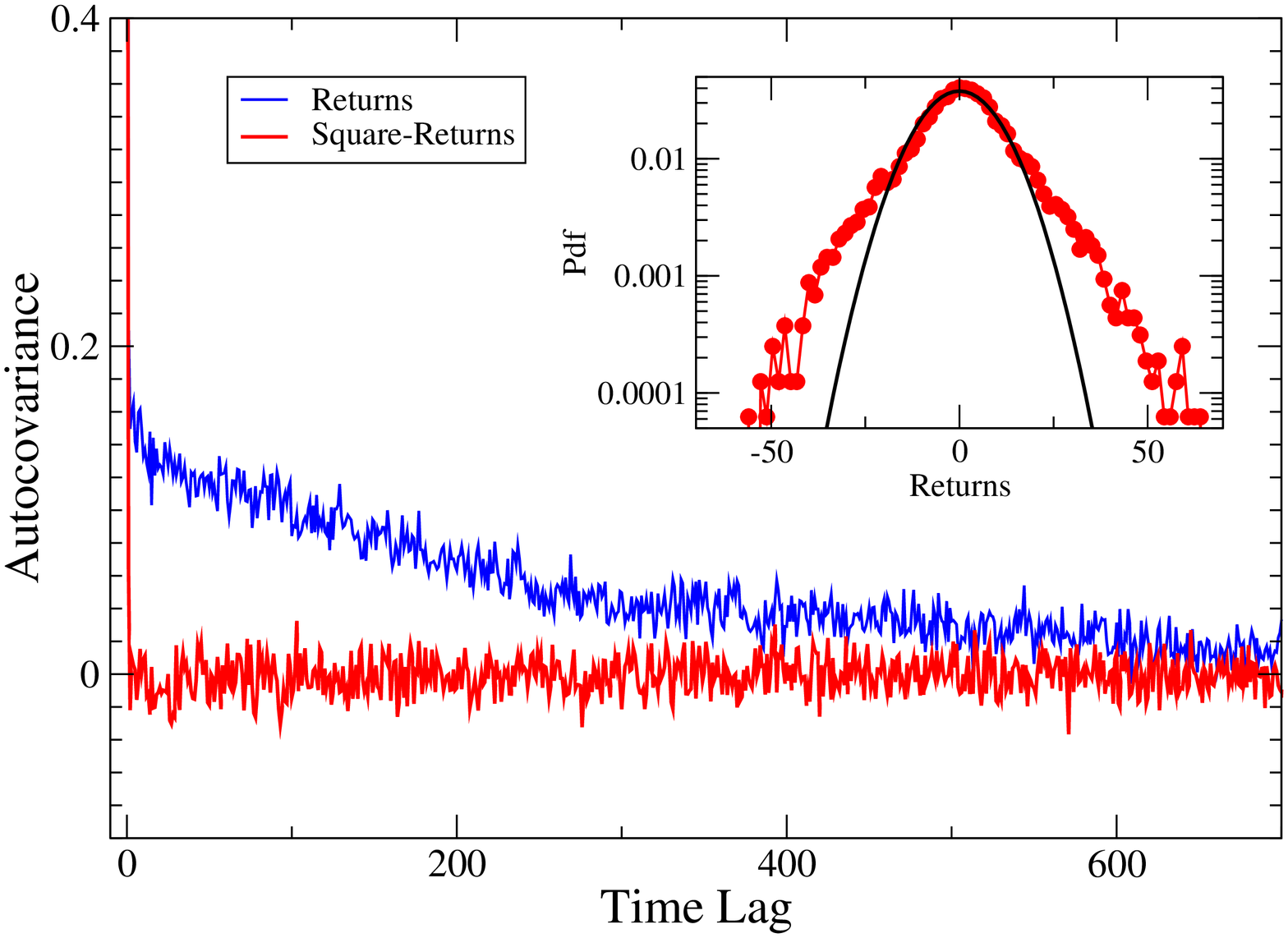}
\caption{We can adjust the parameters of the model in such a way that
the SF appear for any preassigned number of agents. The present case refers
to $N=100$. Also in this case we have omitted the exponential
term of the price behavior in the probability to change strategy.}
\label{fig:34}
\end{figure}

\begin{figure}[h!]
\centering
\includegraphics[scale=0.4]{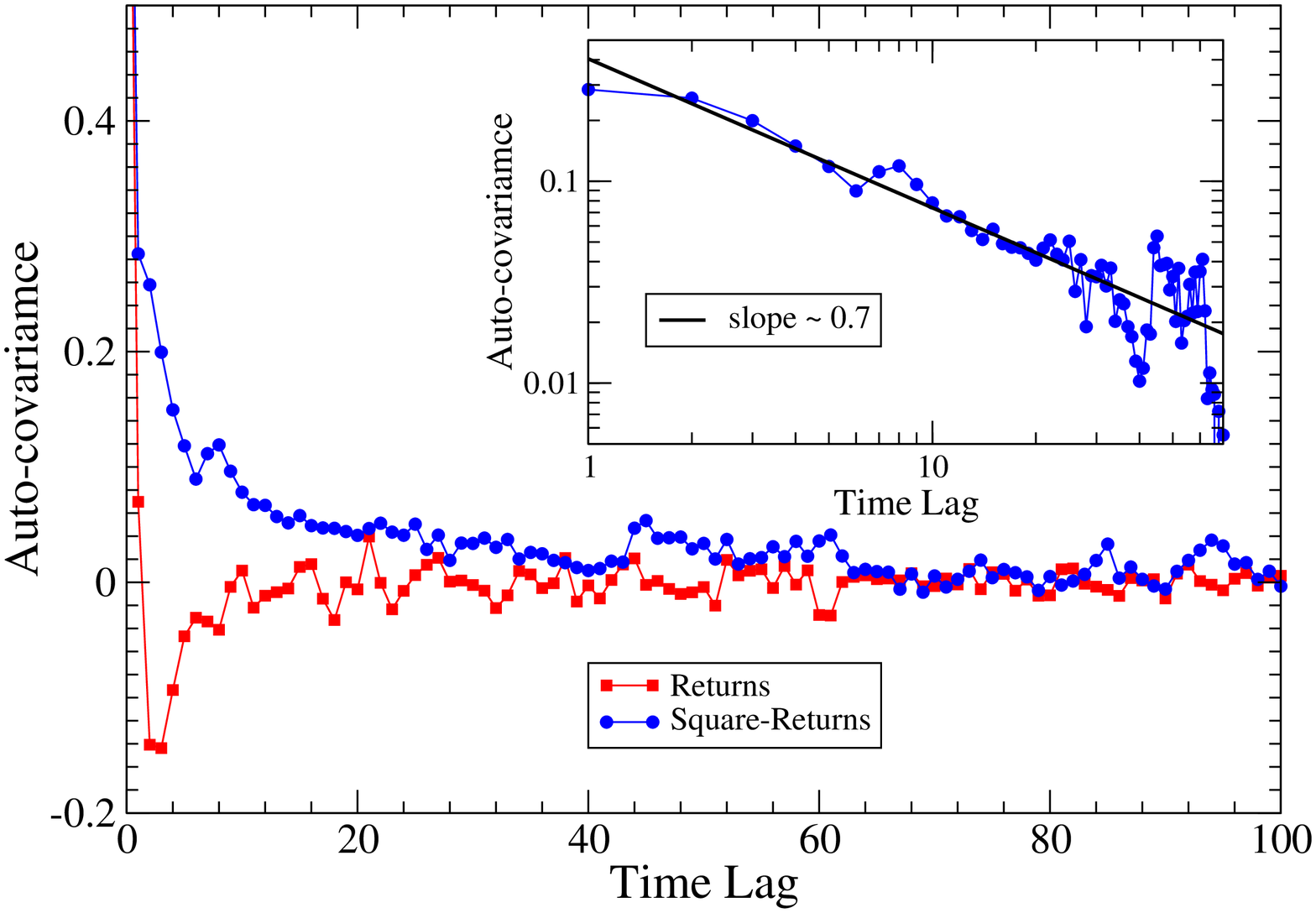}
\includegraphics[scale=0.4]{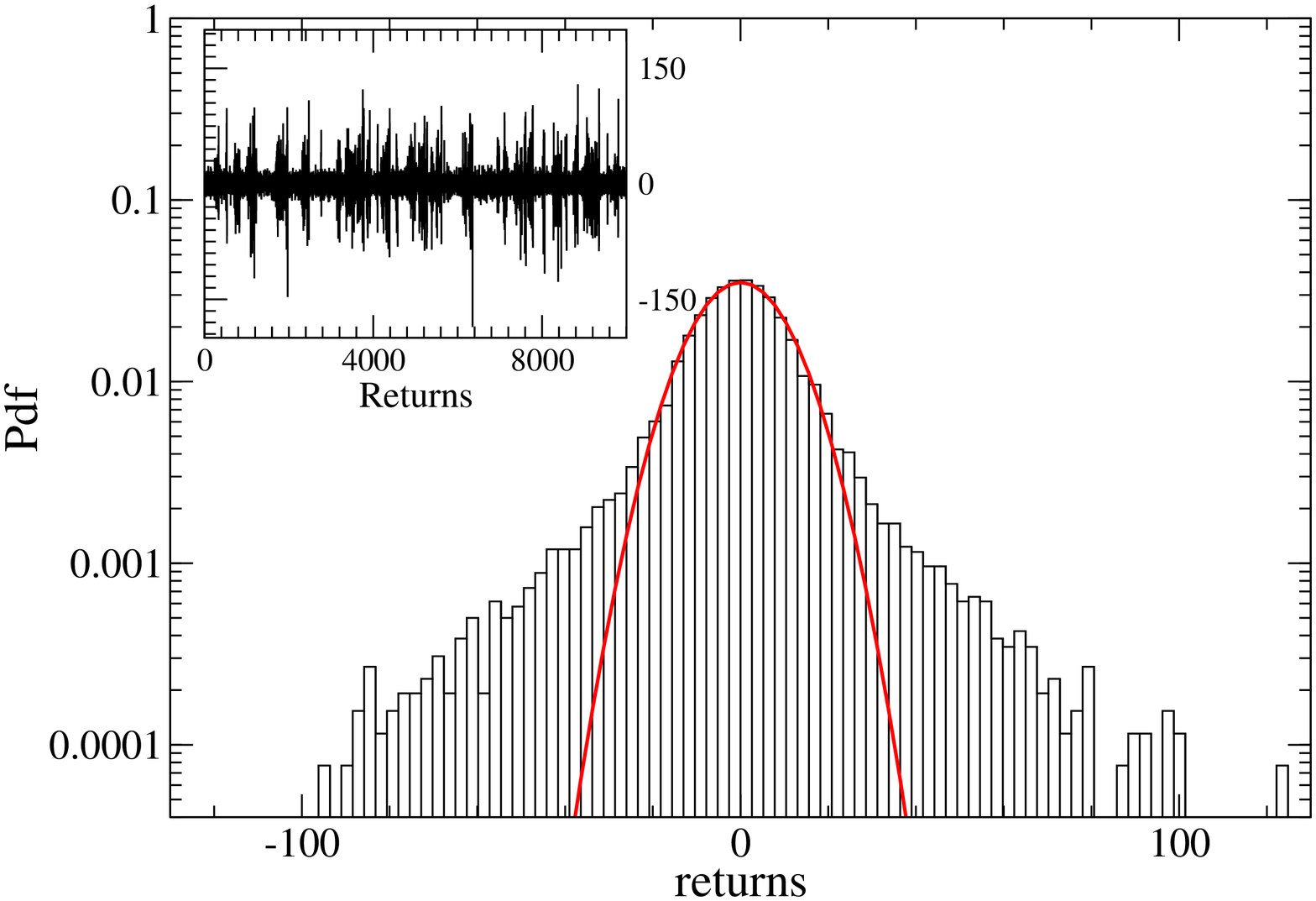}
\caption{Here we show the results for the most complete case we consider
and in this case we adopt $N=500$.
This corresponds to including the exponential price term and introducing
a real heterogeneity in the agents respect to their time horizons
and impact on the market (see text discussion for details).
We can see that the heterogeneity suppresses somewhat the volatility
clustering because a lack of coherence between the agents behavior but
the shape of the volatility time correlation (see insert) in this 
case resembles power law. The exponential price term increases the
size of the fat tails as shown in the lower figure.
}
\label{fig:5}
\end{figure}

In this section we discuss the results of some simulations of the minimal 
model  described
in Sec.~\ref{sec:2}. In particular we will see that this model is 
able to reproduce the main SF of financial markets listed in 
the introduction.
In order to clarify all the elements of the model we are going to discuss increasingly complex cases.\\\\
(a) Single agent\\
Considering that in the simplest model all agents are statistically
identical (no real heterogeneity) we can fix our parameters in 
such a way that even a single stochastic agent can lead
to an interesting dynamics.
This single agent can be chartist or fundamentalist and 
the herding term is not active in this case.
For simplicity we have also neglected the exponential term
related to the price behavior in the transition probabilities.
When the agent switches her behavior from chartist to fundamentalist, 
the market dynamics
is in turn given by Eq. (\ref{eq:0}) or (\ref{eq:2}).
If the agent is chartists she follows the market trend and creates 
bubbles, in this case the
price fluctuations are larger with respect to a simple random walk.
Otherwise if she is fundamentalist the price is driven towards the 
fundamental price and the
fluctuations are smaller than the random walk ones. 
In this case the price is not diffusive
and it remains almost constant, oscillating around the fundamental price $p_f$.
In Fig.~\ref{fig:2} we show the results of the simulation for 
the one-agent model.
We can observe that the price dynamics displays local bubbles corresponding 
to periods in which the agent is a chartist. 
The price is instead almost constant when the
agent is fundamentalist. Also we can observe periods of high or low 
volatility depending
on the strategy of the agent.
This dynamics clearly leads to fat tails in the distribution
of price increments and also to a certain volatility
clustering that certainly in this case is not due
to the herding dynamics.
This simple example shows that, once we have a full control
of the model parameters, we can trace and reproduce some
SF even with an extremely minimal model.\\\\
(b) Many statistically equivalent agents $(N=100)$\\
We now consider the more realistic dynamics with a larger 
number of agents.
In principle we can tune the parameters to
obtain the SF for any preassigned value of $N$.
For example 
in Fig.~\ref{fig:34} (upper)
we show the dynamics of the case $N=100$ still without
the exponential price term. Also in this case periods of 
high or low volatility
correspond to regions in which chartist or fundamentalist agents dominate.
We have analyzed the SF for this model and both fat tails and volatility
clustering are shown in Fig.~\ref{fig:34} (lower). 
The square price-fluctuations shows a positive autocorrelation 
which (unlike real data) 
decays exponentially. As we have discussed this depends
on the fact that this model has a single characteristic time scale.\\\\
(c) Many heterogeneous agents\\
The limitation of a single time scale can be easily removed by 
introducing a real 
heterogeneity in the time scales of the agents' strategies.
In particular we have introduced a distribution of values for the 
parameter $M$ which is the
number of steps agents consider for the estimation of the moving 
average of the price. 
We adopt a distribution of values also for the parameter $b$ 
which gives the strength of the action of 
chartists.
In this case we also introduce the exponential price term in the rate 
equations in order to have the most complete version of our model.
In Fig.~\ref{fig:5} we report the SF
corresponding to a uniform $M$ distribution between $10$ and $50$ time
steps
and a uniform $b$ distribution between $0$ and $2$.
\\
The comparison of these results with those 
of the simplified model discussed before permits to 
draw the following conclusions:\\
- The exponential term related to the price behavior
in the transition probabilities enhances the bubbles and crashes 
corresponding to the chartists action and it amplifies
the fat tail phenomenon as shown by comparing Fig.~\ref{fig:5} (lower)
to Fig.~\ref{fig:34} (lower) .This is a qualitative argument because
the two models are actually different and a precise comparison 
keeping the same effective parameters is not possible.
\\
- The real heterogeneity of the agents time scales produces in
this case a distribution of transition rates which leads to 
a quasi-power law behavior for the volatility correlations as
shown in the inset of Fig.~\ref{fig:5} (upper).
On the other hand this different time scales have the effect of
decorrelating the agents dynamics and the overall amplitude of the volatility
is reduced.\\
-The short time anti-correlation for the returns (Fig.~\ref{fig:5} upper), can be related
to the long time predominance of fundamentalist and will be discussed in detail
in paper II~\cite{paperoII}.

\clearpage

\section{Microscopic origin of the Stylized Facts}

As pointed out pointed out before, the intermittent behavior that characterizes the 
oscillations of
agents' strategies is crucial to generate the Stylized Facts. 
One must choose a certain temporal 
window \textit{and} a correct value of $N$ to obtain the ``right'' amount of 
fluctuations (the other parameters being fixed).  In fact, while a low value 
of $N$ produces too many fluctuations, a high value of $N$ will prevent the formation 
of the chartists bubble at all. Therefore the Stylized Facts in our model can be regarded 
as finite size effects, that is, they disappear in the thermodynamic limit.
Given the general nature of these results we may conjecture that it should apply
toa broad class of agents model and it is not a special property
of our specific model.
\\
This has important consequences for both data analysis and for the 
structure of the models 
proposed to investigate the origin of Stylized Facts. The fact that the exponents of the 
power laws characterizing financial markets don't seem to be universal 
\cite{ramac}
can be 
easily explained in this framework. 
The apparent power laws observed in many data find a natural explanation
in the presence of a distribution of agents' strategies in terms
of their time horizons and strength. Concerning universality, even if this 
is not strictly present in this model, it is reasonable to expect a certain 
similarity in all markets. This is due to the fact that the key elements
to generate the SF are of very general nature as we are going to explain in
the following discussion.\\
The simplicity of the model permits to interpret the origin of Stylized Facts 
directly from the agents' strategies in the market. To this purpose it will 
be useful to define the concept of \textit{effective action}. In our model 
agents not only decide between selling and buying, they operate in the market proportionally 
to the \textit{signal} $p-p_f$ or $p-p_M$, selling or buying quantities
which are proportional to this signal.
Besides this, while in Statistical Mechanics
the study of the dynamics of a model is usually done with a fixed value of $N$
(and eventually infinite), nothing in financial markets permits us to justify this 
assumption. An agent can either enter or exit from the market on the basis of various 
considerations, and in addition she can varies the volume of the exchanges in the market. 
In summary, we must consider the effective action present in the market as the sum of these 
two effects, the fact that an agent can operate with different volumes and the fact that 
agents can enter or exit from the market.\\
Let us now suppose that for some reason at time $t$ there is a price fluctuation  
(in any direction) $\Delta p$. Following our line of reasoning, this will produce 
an increase of the effective action, because both fundamentalists and chartists will 
see a signal in the market, and this action will produce more price fluctuations, and so on. 
So action leads to more action. On the contrary, if the market doesn't show 
opportunities to be profitable, agents will be discouraged from operating and so 
periods of low fluctuations will be followed by periods of low fluctuations. 
This mechanism resembles the GARCH process, in which the volatility at time $t+1$ 
depends on the volatility at time $t$ and also on the return at time $t$~\cite{garch}:
\begin{equation}
 \sigma(t+1)=f(\sigma(t);\Delta p(t))
\end{equation}
We recall that the GARCH process was originally proposed as a phenomenological scheme to 
reproduce the phenomena of volatility clustering and fat tails.
In our case we propose a microscopic interpretation of something
similar to the GARCH process which, however, is now related to the
specific agents dynamics.
\\
On the other hand our model doesn't show any appreciable linear correlation between 
price increments, even if the arbitrage condition is not explicitly implemented
in  the model. 
This can be understood in the following way. Consider a price increment (with sign) 
$\Delta p$ at time $t$. The next price increment  will depend not only from the 
previous fluctuation, as it is for volatility, but also on all the other variables 
of the system, 
like the number of chartists, the specific values of $p_M$ and $p_f$ and so on.
Schematically we can write:
\begin{equation}
 \Delta p(t+1)=f(\sigma;N_c;N_f;p_M;p_f).
\end{equation}
All these additional variables are in general not correlated in a direct
way with the price fluctuations, so they lead to a decorrelation
of the price increments, as is, instead, in the case of volatility.\\
This line of reasoning explains qualitatively the presence of volatility clustering and 
the absence of linear correlations in financial markets. We believe that these
considerations are of general nature and therefore they should be valid for all
markets and models.
Within our class of models, however, this general behavior does not reach the
status of {\itshape universal behavior} in the sense of Statistical Physics.
\clearpage

\section{Self Organized Intermittency}

\begin{figure}[h!]
\centering
\includegraphics[scale=0.5]{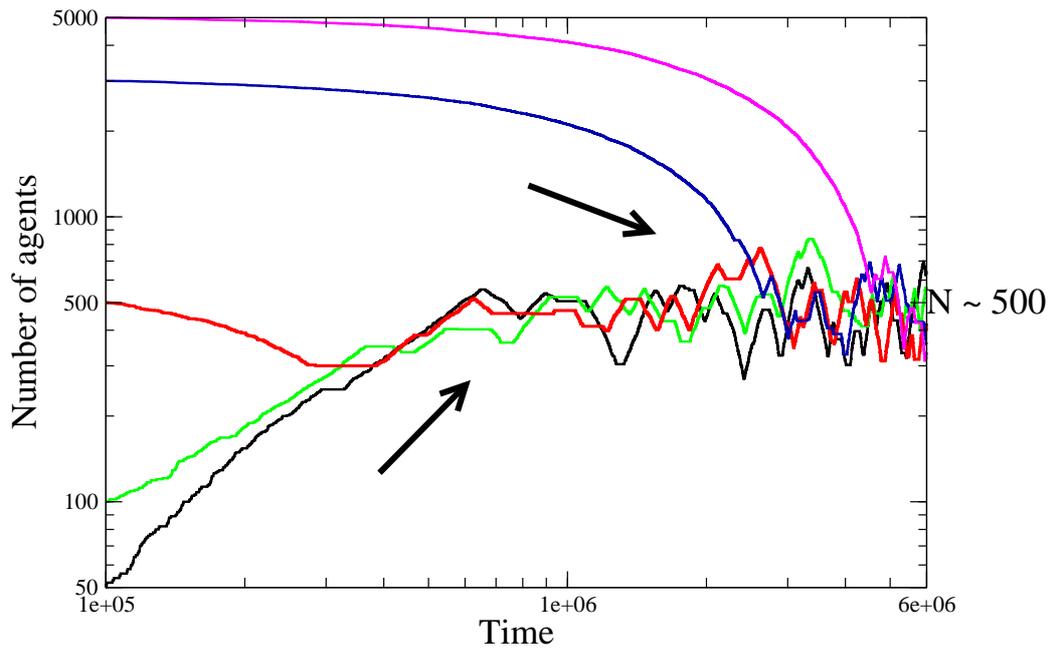}
\caption{Self-Organized Internittency. We can observed the self-organized nature of the dynamics toward to the quasi-critical state corresponding to existence of Stylised Facts. We show five cases, characterized by different initial conditions, $N(t=0)=$ 5000, 3000, 500, 100 and 50. If the number of the agents in the market is too large $(N=5000, 3000)$ the price will be stable and essentially driven by fundamentalists. In this case the market is not profitable and agents will not operate. So $N$ will decrease. On the other hand a too small number of agents $(N=50, 100)$ will lead to large fluctuations and will attract more agents, increasing the value of $N$. In all cases the system converges to an intermediate situation $(N\approx 500$ in this case) which corresponds to the intermittent behavior which leads to Stylized Facts. Note that this regime is not real critical and universal in the sense of Statistical Physics.}
\label{5soc}
\end{figure}

In the previous section we have seen that our model is able to 
generate the SF of financial markets and it is possible to control their 
origin and nature in great detail.
However, in our model, as in most of the models in the literature,
these SF occur only in a very specific and limited region of
the models' parameters.
This is a problem which is seldom discussed in the literature
but in our opinion poses a vary basic question: why the market dynamics
evolves spontaneously, or self-organizes, in the specific region of parameters which
corresponds to the SF? In this section we are going to discuss in detail a possible
answer to this fundamental question.\\
In usual Critical Phenomena of Statistical Physics 
there is  a basic difference between the
parameters of the model (usually called coupling constants) and 
the number $N$ of elements considered.
Models with different coupling constants, but belonging to the same universality class,
evolve  towards the same critical properties in the asymptotic limit of very large $N$ 
and very large time.
This situation is called universality and it is often present
in equilibrium critical phenomena in which the critical region requires
an external fine tuning of various parameters. 
This is a  typical situation of competition between order and disorder
which occurs at the critical point.\\
Self-organization instead occurs  in a vast class of models 
characterized by a non linear dissipative dynamics far from equilibrium.
Popular examples of these self organized models
are
the fractal growth models like
of Diffusion Limited Aggregation (DLA), the Dielectric Breakdown Model (DBM)
and the Sandpile Model~\cite{DLA,LPdbm,SOCbak,jensen}.
In these models the nonlinear dynamics
drives spontaneously system towards a critical state.
This can occur from a variety of initial configurations and parameters
and define a state that is always the same , the critical one.
For these reasons this phenomenon has been named Self-Organized Criticality (SOC)~\cite{bak}.
It is important to note that this self-organization is also an asymptotic
phenomenon, in the sense that it occurs in the limit of large N and large time.
\\
For financial markets the possible analogy with SOC
phenomena 
is very tempting. However we are going to see that there are basic fundamental differences
with the above concepts which require the developments of a different theoretical framework.
In this section we are going to introduce the first step along this line.
\\
A very important observation is the fact that the SF appear only for a specific value of 
the number of agents $N$. This result may appear as problematic in view of the above
discussion about self-organization.
A finite value of $N$ cannot lead to universality
and so the presence of the SF in  virtually all markets
(with very different number of agents)
appears rather mysterious.
In this 
perspective a situation in which the SF are generated only in the limit 
of large $N$ would have appear more natural. 
Also our model shows clearly that 
the intermittent behavior corresponding to the SF must necessary occur for a finite value 
of $N$. It is easy to realise that this conclusion has a general nature with respect to the 
four essential ingredients of our class of models. An additional problem, in trying to 
explain the self-organization of the SF, 
comes from the fact, that in addition to N, the model contains several other parameters. 
So, even if one would be able to produce the SF in the large $N$ limit, the self organization 
with respect to the other parameters would remain unclear.\\
From our studies we are going to propose a conceptually different
mechanism of the 
self-organization phenomenon. Consider a certain market with its characteristic 
parameters. This values will govern the rate equation of the agents dynamics as
described by Eqs.~\ref{eq:3}and ~\ref{eq:4}
and in Fig.~\ref{flsgfd}.
For any set of these parameters there will be a characteristic value of $N$ 
which separates the regions with large fluctuations with the ones with small fluctuations, 
as shown for example in Fig.~\ref{flsgfd}
The SF appears precisely around this finite value of $N$, which corresponds to the right amount
of intermittency.
So our basic question about the SO is now transformed in a question related to the number of 
agents. Why the system should have precisely this number of agents acting on the market?
\\
This question can now be answered in the following way.
We have seen that if $N$ is very large the system gets locked in the fundamentalist state
and this leads to a very stable dynamics of the price.
Such a situation will produce very small signals for the agent strategies. 
If we assume that 
an agent operates in an certain market only if her signal is larger than a minimum threshold,
it is easy to realise that in the stable price situation, corresponding to large $N$,
the number of active agents will decrease.
On the other hand in the case of very few agents, the price dynamics shows very
high fluctuations~\cite{paperNP}
and this produces large signals for the agents and will attract  
more agents in the market.
Mathematically this concept can be implemented by
introducing a trashold to the price fluctuations which 
corresponds to the decision of agents to be
active or not active in the particular market.
We have introduced this threshold in our model and a typical result
is shown in Fig.~\ref{5soc}
We can see that the variable number of agents evolves spontaneously
towards the characteristic value of $N$ corresponding to the
intermittent state and the SF.
Since this state is not precisely critical in the sense of Statistical Physics
we propose to call this phenomenon {\itshape Self 0organized intermittency } (SOI).
This new vision has important implications in various directions. On one hand it seems 
to be a reasonably natural and robust explanation of the self-organization towards the 
quasi-critical state, on the other hand the quasi-critical state leading to the SF
 is unavoidably linked to finite size effect. A first consequence is that in the limit 
of large $N$, or 
in the limit of large time, the SF disappear. A fact which seems to be indeed reproduced 
by price time series~\cite{ramac}.
Another consequence is that genuine critical exponents and universality can not be expected 
in this framework. The fact that several datasets can be fitted by power laws is however 
not surprising. Since different agents may have different time horizons in 
their trading strategies, one has a superposition of finite size effects which
 may appear as a power law in some range. This perspective implies that the discrepancy 
in the effective exponents for different dataset and for different time scales \cite{ramac}  
may be a genuine intrinsic property instead of spurious effect due to the incompleteness of 
the data.\\
A key element of the present scheme is the nonstationarity of the system with respect
to the  value of $N$ which represent a major departure from usual critical
phenomena. On the other hand this non stationarity appears to be a very important
element in real financial markets.
\clearpage
\section{Conclusions}

Our starting point has been a detailed analysis of the LM model first
in terms of the role played by the total number of players $N$ and also
on the study of the stability of the SF with respect to the variation
of the other parameters.  A particular puzzling point was the fact that
the SF are present only for a finite value of $N$ $(N=500)$ in the LM
model) and not in the asymptotic limit $N \rightarrow \infty$.
Starting from this consideration our aim has been to introduce a
minimal ABM based on F and C which would permit to clarify these
points and eventually also to discuss the self-organization.  A basic technical
simplification is the description of the chartists in terms of the
newly introduced effective potential model.  This and other similar
simplifications permit us to analyze in detail the mathematical
properties of the model.  In order to achieve the deepest
understanding also analytically of the dynamics we have focused for
the moment on the linear returns.  In paper II~\cite{paperoII}
we are going to consider also the extension to logarithmic returns
but in the limit of small fluctuations the two approach essentially
coincide.
The main results of our model are:\\
- Detailed understanding of the origin of the SF with respect to the
microscopic
dynamics of the agents.\\
- Demonstration that in this class of model the SF correspond to
finite size effects and not to universal critical exponents. This
finite size effect, however, can be active at
different time scales.\\
- Bubbles of chartists can be triggered spontaneously by a
multiplicative cascade which can originate from tiny random
fluctuations.  This situation resembles in part the avalanches of the
Sandpile Model in Statistical Physics~\cite{bak}.\\
- We have shown the importance of the non stationarity in the dynamics
of the number of active agents $N$ and we introduced a characteristic
threshold to decide when
an agent  can enter or exit from the market.\\
- This threshold and the relative non stationarity are proposed to
represent the key element in the  self-organization mechanism. This self
organization, however, leads to an intermittency related to finite
size effects. For this reasons we define it as 
Self-Organized-Intermittency (SOI).
\\
Starting from the minimal model introduced in this paper and
considering that one can obtain a detailed microscopic understanding
of its dynamics, it is easy to identify a number of realistic variances
which can be introduced as generalizations of the model. In future
works we will consider this variantes. However, the present model was
aimed at a different target.  The idea was to define the minimal set
of elements which could lead to the SF and to the SO phenomenon.

\section*{Acknowledgments}
We are grateful to  Simone Alfarano,  Guido Caldarelli, Alessio Del Re, Doyne Farmer, Fabrizio Lillo, Thomas Lux, Rosario Mantegna, Miguel Virasoro, and Constantino Tsallis for interesting discussions.

\clearpage
\bibliographystyle{hunsrt} 
\bibliography{merging}

\end{document}